\documentclass[aps,preprint,showpacs,preprintnumbers,amsmath,amssymb,superscriptaddress]{revtex4}

\usepackage{graphicx}
\usepackage{dcolumn}
\usepackage{bm}
\usepackage{color}

\newcommand{\vecr}{\vec{r}}

\newcommand{\vb}{\vec{b}}

\newcommand{\gammaa}{\gamma_{\rm rc}}

\newcommand{\psitrap}{\psi_{\rm out}}
\newcommand{\hatpsitrap}{\hat{\psi}_{\rm out}}
\newcommand{\psireact}{\psi_{\rm rc}}
\newcommand{\hatpsireact}{\hat{\psi}_{\rm rc}}

\begin{document}
%
\title{
Dispersive photoluminescence decay by geminate recombination in amorphous semiconductors
}

\author{
Kazuhiko Seki}
\email{k-seki@aist.go.jp}
\affiliation{
National Institute of Advanced Industrial Science and Technology (AIST)\\
AIST Tsukuba Central 5, Higashi 1-1-1, Tsukuba, Ibaraki, Japan, 305-8565
}
\author{K. Murayama}
\email{
murayama@phys.chs.nihon-u.ac.jp}
\affiliation{
College of Humanities and Sciences, Nihon University, Tokyo 156-8550, Japan
}
\author{M. Tachiya
}
\email{m.tachiya@aist.go.jp}
\affiliation{
National Institute of Advanced Industrial Science and Technology (AIST)\\
AIST Tsukuba Central 5, Higashi 1-1-1, Tsukuba, Ibaraki, Japan, 305-8565
}
\date{\today}
\begin{abstract}

The photoluminescence decay in amorphous semiconductors is described by power law $t^{-\delta}$ at long times.
The power-law decay of photoluminescence at long times is commonly observed but recent experiments have revealed 
that the exponent, $\delta \sim 1.2-1.3$, is smaller than the value $1.5$ predicted 
from a geminate recombination model assuming normal diffusion. 
Transient currents observed in the time-of-flight experiments are highly dispersive characterized by 
the disorder parameter $\alpha$ smaller than $1$.
Geminate recombination rate should be influenced by the dispersive transport of charge carriers. 
In this paper we derive the simple relation, $\delta = 1+ \alpha/2 $.
Not only the exponent but also the amplitude of the decay calculated in this study is consistent with 
 measured photoluminescence in a-Si:H. 
\end{abstract}
\maketitle

\newpage
\setcounter{equation}{0}
\section{Introduction}
\vspace{0.5cm}

The photoluminescence in amorphous semiconductors consists of a major band and weaker bands of longer wavelength attributed 
to dangling bonds \cite{Street}. 
The main photoluminescence band arises from the radiative recombination of electrons and holes both trapped in band tail states \cite{Street}. 
Our interest is in this main band excited with relatively weak light pulses at the absorption edge. 
Photoluminescence decay 
is described by power law $t^{-\delta}$ at long times above a certain temperature \cite{Street,NHS,Oheda,MN85AsS,MN85,Murayama2002,Ando,FDAMurayama}. 
For sufficiently law excitation intensity 
a geminate recombination model for photoluminescence decay has been developed, 
which predicts a $t^{-3/2}$ decay of luminescence at long times \cite{NHS,HNS}. 
In the model the $t^{-3/2}$ behavior at long times is attributed to normal diffusion of a geminate pair. 
Predicted $t^{-3/2}$ decay of luminescence at long times is confirmed by early experiments, 
where high temperature data is different from the low temperature decay for which diffusion is negligible \cite{Street,NHS}. 
At high temperatures electrons have enough thermal energy to conduct hopping transport in the band tails \cite{Street,NHS,HNS}.
It is widely recognized that a hole is trapped at a deep localized 
state at very short times after the pulse excitation  \cite{Street}. 
The power-law decay of photoluminescence at long times is commonly observed but recent experiments have revealed 
that the exponent is smaller than the predicted value, $\delta \sim 1.2-1.3 < 1.5$ \cite{Oheda,MN85AsS,MN85,Murayama2002,Ando}. 
The exponent increases with increasing temperature. 
In other words, the deviation of the exponent from $1.5$ decreases by raising temperature. 
Such deviation is interpreted in terms of deviation of motion of charges from the normal diffusion \cite{Oheda,MN85AsS,MN85,Murayama2002,Ando}.

Indeed, at temperatures at which power law of the photoluminescence decay is observed with $\delta$ smaller than $1.5$ 
transient currents observed in the time-of-flight experiments are highly dispersive and completely different from those 
obtained from normal diffusion \cite{SM,Scher,TiedjePRL,Tiedje}.
The shape of the transit current, $I$, is featureless in the conventional units $I$ versus $t$ \cite{SM,Scher,TiedjePRL,Tiedje}.
The transit time, $t_{\rm tr}$, is obtained only from the $\log I$ versus $\log t$ traces, 
because many time scales coexist for electron hopping between localized sites in the band tail states \cite{Scher}. 
The shape of the transit current is derived by a continuous time random walk model 
where the hopping time distribution function has an algebraic asymptotic form \cite{SM,Scher,Schnorer,Jakobs} 
$\psi (t) \sim \gamma_{\rm r}^{- \alpha} t^{-(1+\alpha)}$ , where 
$\gamma_{\rm r}$ is a carrier jump frequency  
and $\alpha$ is a constant in a range of $0 < \alpha < 1$.
From the current traces 
the disorder parameter, $\alpha$, is typically obtained as $\alpha \sim 0.6$ \cite{SM,Scher,TiedjePRL,Tiedje,MurayamaMC5}. 
It implies that  
the displacement, $r$, of a charge carrier obeys 
$\langle r^2 (t) \rangle \propto t^\alpha$. 
Therefore, electron transport cannot be described by normal diffusion for which 
$\langle r^2 (t) \rangle \propto t$.

Geminate recombination rate should be influenced by dispersive transport of charge carriers 
\cite{Oheda,MN85AsS,MN85,Murayama2002,Ando}.  
The fact is widely recognized but a simple relation between $\delta$ and $\alpha$ is not established. 
In this paper we derive a simple relation between the long time exponent of photoluminescence decay, $\delta$, 
and the disorder parameter, $\alpha$. 
The relation is confirmed by experimental data of a-Si:H taken at various temperatures.  
In the experiments, 
the samples of a-Si:H were prepared by a glow discharge method. 
The sample held in helium gas was excited with $10$ nsec light pulses of energy around $2.0$ eV at 
the absorption edge in a-Si:H.
The photoluminescence main band is observed at the peak energy $1.22-1.35$ eV. 
We also obtain a theoretical relation between 
amplitude of the power law decay of photoluminescence and the disorder parameter, $\alpha$. 
The amplitude decreases due to escape of an electron-hole pair from recombination. 
The result is also confirmed by measured photoluminescence data.

\setcounter{equation}{0}
\section{Photoluminescence}
\vspace{0.5cm}

Amorphous semiconductors have localized states in band tails \cite{Street}. 
The localized states are randomly distributed and their electrostatic energies are also distributed. 
In a-Si:H electron-hole pairs are created by light irradiation. 
Holes are quickly trapped and electrons execute hopping-random walks among localized band tail states \cite{Street}. 
Radiative recombination of geminate electron-hole pairs is observed as photoluminescence 
whose decay curve is independent of the excitation intensity in a range from $5 \mbox{[}n\mbox{J}/\mbox{cm}^2 \mbox{]}$ to
$500 \mbox{[} \mu \mbox{J}/\mbox{cm}^2 \mbox{]}$ \cite{MN85}. 
Intensity of photoluminescence is proportional to the recombination rate 
which is given by the decay rate of survival probability of electrons. 
Long time decay of photoluminescence is controlled by diffusion of electrons to the sites where holes are located. 
Motion of electrons due to hopping-random walks has been measured by 
transient current in the time-of-flight experiment \cite{SM,Scher,TiedjePRL,Tiedje}. 
The electron transport is sub-diffusive where the mean square displacement of electrons obeys 
\begin{eqnarray}
\langle r(t)^2 \rangle \propto t^{\alpha} .
\end{eqnarray}
The initial distance of electron-hole pairs may have spatial distribution even immediately after photo-excitation. 
However, the long time asymptotic kinetics of photoluminescence decay 
is insensitive to the initial distance, $r_0$. 
The distribution of initial distance immediately after photo-excitation is important 
to analyze short time kinetics of photoluminescence. 
In this paper we focus on the long time asymptotic kinetics. 
As written before, intensity of photoluminescence is proportional to the recombination rate. 
In the limit of infinitely fast intrinsic recombination rate (recombination rate at an encounter distance)
compared to the thermal jump frequency of electrons 
the recombination reaction rate, $R_\infty (\vecr_0, t)$, is nothing but the probability distribution of first 
arrival at the encounter distance $R$; namely, the first-passage time density. 
The transition probability or the conditional probability 
$g \left( \vecr_0, t \right)$ of finding an electron at the encounter distance at time $t$ when it starts initially at site $\vecr_0$ 
in the system with no reaction satisfies the integral equation \cite{Weissbook,Haus},
\begin{eqnarray}
g \left( \vecr_0, t \right) = \int_0^t d t_1 g \left(R, t -t_1 \right) R_\infty \left( \vecr_0, t_1 \right) 
\label{chainR}
\end{eqnarray}
because random walker at the encounter distance at time $t$ must have been there for the first time at time $t_1 \leq t$. 
After Laplace transformation of Eq. (\ref{chainR}) 
$\hat{R}_\infty \left( \vecr_0, s \right)$ is obtained as 
\begin{eqnarray}
\hat{R}_\infty \left( \vecr_0, s \right) = 
\frac{\displaystyle \hat{g} \left( \vecr_0, s \right)}{\displaystyle \hat{g} \left( R, s \right)} .
\label{Rinf}
\end{eqnarray}
In experiments the intrinsic recombination rate is not infinitely large but it competes with jump frequency of electrons. 
By increasing temperature the luminescence quantum efficiency decreases, 
which indicates that more electrons escape recombination due to 
increase of hopping frequency. 
The effect is strong when the hopping frequency competes with the intrinsic recombination rate. 
When recombination rate competes with jump frequency of electrons,  
recombination rate, $R \left( \vecr_0, t \right)$, 
is obtained from $R_\infty \left( \vecr_0, t \right)$ by the method of Pedersen \cite{Pedersen,Tachiya}.

First we note that the probability that a particle which starts at $r_0$ will visit the spherical shell of 
 radius $R$ during the time interval between $t$ and $t + dt$ is equal to the recombination rate for the perfectly absorbing boundary condition. 
Two types of waiting time distribution is defined inside the reaction zone; 
one is the waiting time distribution function of making a jump without reaction, $\psitrap (t)$, 
and the other is the waiting time distribution function of reaction, $\psireact (t)$.
The waiting time distribution of making a jump at the encounter distance is given 
in terms of that in the absence of reaction, $\psi(t)$, as, 
\begin{eqnarray}
\psitrap (t) =  \psi (t) \exp \left( - \gammaa t \right) .
\label{wtdout}
\end{eqnarray}
The waiting time distribution of reaction is defined as 
the probability that the particle which is initially at a site in 
the encounter distance will undergo reaction without making a jump at time $t$. 
It is given by the 
reaction rate constant,  $\gammaa$, multiplied by the 
remaining probability of particles at the site in the reaction zone, which decays 
either by jump motion or reaction,  
\begin{eqnarray}
\psireact (t) &=& \gammaa \exp \left( - \gammaa t \right)
\int_t^\infty d\,t_1 \psi(t_1) .
\label{psireaction}
\end{eqnarray}
Some particles may recombine at the first visit to the encounter distance $R$. 
Other fraction of particles escape recombination at the first encounter and make a jump to the spherical 
sphere of radius $R+b$ with $b$ being the jump length.
Some particles may recombine at the second encounter after jumping from the sphere of radius $R+b$. 
The recombination rate $R \left( \vecr_0, t \right)$ should be the sum of the contributions from the first encounter, 
second encounter and so on, 
\begin{eqnarray}
\lefteqn{R \left( \vecr_0, t \right) = 
\int_0^t d\, t_1 \psireact \left( t - t_1 \right) R_\infty \left( \vecr_0, t_1 \right)
+ } 
\nonumber \\
&&
\int_0^t d\, t_1 \int_0^{t_1} d\, t_2 \int_0^{t_2} d\, t_3 
\psireact \left( t - t_1 \right) 
R_\infty \left( R+b, t_1 - t_2 \right) 
\psitrap \left(t_2 - t_3 \right) R_\infty \left( \vecr_0 , t_3 \right) + \cdots .
\label{PedersenR}
\end{eqnarray}
By Laplace transformation Eq. (\ref{PedersenR}) becomes \cite{SekiOctober}, 
\begin{eqnarray}
\hat{R} \left( \vecr_0, s \right) &=& 
\hat{R}_\infty \left( \vecr_0 , s \right)
\frac{\hatpsireact (s)}
{1 - \hatpsitrap (s) \hat{R}_\infty \left( R+b, s \right)} .
\label{Pedersen}
\end{eqnarray}
Therefore, $\hat{R} \left( \vecr_0, s \right)$ is obtained from $\hat{R}_\infty \left( \vecr_0 , s \right)$, 
which is expressed in terms of the transition probability $\hat{g} \left( \vecr_0, s \right)$ as Eq. (\ref{Rinf}). 
When migration of charges is due to thermal transition between trap states 
and is described by fractional Brownian motion,   
the transition probability is given by \cite{Metzler2004}, 
\begin{eqnarray}
\hat{g} \left( \vecr_0, s \right) = \frac{\displaystyle \exp \left[- \frac{r_0 - R}{\sqrt{\displaystyle D_\alpha s^{-\alpha}}} \right]}
{4 \pi D_\alpha r_0 s^{1-\alpha} \left( 1 + \frac{\displaystyle R}{\displaystyle \sqrt{\displaystyle D_\alpha s^{-\alpha}}} \right)} ,
\label{gs}
\end{eqnarray}
where the perfectly reflecting boundary condition is imposed at $r=R$.
When the fractional Brownian motion is derived from the continuous time random walk with jump length $b$ 
and jump frequency $\gamma_{\rm r}$,   
a generalized diffusion coefficient is derived \cite{BMK,SekiJuly}, 
\begin{eqnarray}
D_\alpha &\equiv& \frac{\sin \pi \alpha}{2 \pi \alpha} \gamma_{\rm r}^{\alpha} b^2 ,
\label{gD}
\end{eqnarray}
from the precise form of the waiting time distribution function of the release 
from the trap in the absence of recombination reaction \cite{Schnorer,Jakobs}, 
which is written after Laplace transformation,
\begin{eqnarray}
\hat{\psi }(s) \sim 1-\frac{\pi \alpha}{\sin \pi \alpha} \left( \frac{s}{\gamma_{\rm r}} \right)^\alpha .
\end{eqnarray}
By substituting Eq. (\ref{gs}) into Eq. (\ref{Rinf}) we get $\hat{R}_\infty \left( \vecr_0 , s \right)$, 
which is related with $\hat{R} \left( \vecr_0, s \right)$ through Eq. (\ref{Pedersen}). 
The recombination rate in Laplace domain is derived as, 
\begin{eqnarray}
\hat{R} \left( \vecr_0, s \right) = \frac{R}{\displaystyle r_0} 
\exp \left[- \frac{r_0 - R}{\sqrt{\displaystyle D_\alpha s^{-\alpha}}} \right] 
\frac{1}{\displaystyle 1 + 4 \pi R \frac{\displaystyle D_\alpha}{\displaystyle k_\alpha}
\left( 1+ \frac{R}{\displaystyle \sqrt{\displaystyle D_\alpha s^{-\alpha}}} \right)} , 
\label{solL}
\end{eqnarray}
where a generalized intrinsic reaction rate is defined as \cite{SekiJuly,FDASeki}
\begin{eqnarray}
k_\alpha \equiv  
\hatpsireact (0) \frac{\sin \pi \alpha }{\pi \alpha} 
\gamma_{\rm r}^{\alpha} 2 \pi R^2 b
\sim \gamma_{\rm rc}^\alpha 2 \pi R^2 b   .
\label{gk}
\end{eqnarray}
After inverse Laplace transformation the asymptotic photoluminescence decay is obtained as 
\begin{eqnarray}
R \left( \vecr_0, t \right) \sim \frac{\alpha}{2 \Gamma \left( 1 - \alpha/2 \right)} \frac{\displaystyle R}{\displaystyle r_0} 
 \left( \frac{\displaystyle  r_0/ R  -1}{\displaystyle 1+ 4 \pi R D_\alpha /k_\alpha} +
 \frac{\displaystyle 4 \pi R D_\alpha /k_\alpha}{\displaystyle \left( 1+ 4 \pi R D_\alpha /k_\alpha \right)^2} 
\right)
\frac{\displaystyle R}{\displaystyle \sqrt{\displaystyle D_\alpha}} 
\frac{\displaystyle 1}{\displaystyle t^{1+\alpha /2}} . 
\label{asympt}
\end{eqnarray}
According to Eq. (\ref{asympt}) the power of asymptotic photoluminescence decay, $1/t^\delta$, 
is related to the disorder parameter $\alpha$ by 
\begin{equation}
\delta = \frac{\alpha}{2} + 1 , 
\end{equation}
for any strength of the intrinsic recombination rate, $k_\alpha$, 
and the initial distance between the electron and the hole of a geminate pair.  
In Fig. 1 the exponent $\delta$ of the long time power-law decay of photoluminescence 
is shown together with the exponent 
$1+\alpha/2$ calculated from the disorder parameter, $\alpha$, of 
the dispersive transient current, 
which is measured by electron time-of-flight experiments at low electric fields \cite{MurayamaMC5}. 
The two exponents are  consistent, 
which indicates that the long time photoluminescence is controlled by the dispersive transport of electrons.

The intensity of photoluminescence depends on the strength of intrinsic recombination rate and 
the jump frequency of electrons. 
The intensity of photoluminescence decreases by increasing temperature 
because the number of electrons which escape recombination increases. 
We neglect other non-radiative processes 
which are important in samples with a substantial defect density \cite{Street}. 
Obviously, the intensity is also influenced by the initial distance between the electron and the hole of a geminate pair and  
the intrinsic recombination rate. 
We investigate two extremes, separately. 
The first one corresponds to the band-edge excitation, 
where an electron is created in the vicinity of a hole and the intrinsic recombination rate competes with the dispersive transport of electrons. 
The second one corresponds to excitation of electrons to the conduction band 
through which an electron is separated from a hole by large distance and the recombination occurs 
whenever the electron encounters the hole.

In the first case the initial distance between an electron and a hole is small and 
intrinsic recombination rate $k_\alpha$ is small.
An electron may visit the encounter distance many times before recombination. 
In experiments  
the wavelength of incident light is chosen to excite electrons in the tail of band edge 
and electrons are excited close to the immobile holes. 
For simplicity the initial distance of an electron-hole pair is assumed to be 
in the vicinity of the 
encounter distance, $R$
($R+\epsilon$,  $\epsilon \rightarrow 0$)
. 
The recombination rate is obtained from 
the inverse Laplace transform of Eq. (\ref{solL}) as
\begin{eqnarray}
R \left( r_0=R, t \right)  = 
\frac{\displaystyle k_\alpha \sqrt{\displaystyle D_\alpha t^\alpha}}{\displaystyle 4 \pi R r_0 D_\alpha t}
E_{\alpha/2, \alpha/2} 
\left( - 
\frac{\displaystyle 1+ 4 \pi R D_\alpha / k_\alpha}{\displaystyle 4 \pi R D_\alpha / k_\alpha}
\frac{\displaystyle  \sqrt{\displaystyle D_\alpha t^\alpha}}{\displaystyle R}
\right) , 
\label{comps}
\end{eqnarray}
where the generalized Mittag-Leffler function is defined as \cite{Podlubny}
\begin{eqnarray}
E_{a,b} \left( x \right) = \sum_{k=0}^\infty \frac{\displaystyle x^k}{\displaystyle \Gamma \left(a k + b \right)} .
\label{gML}
\end{eqnarray}
In the long time limit the above equation reduces to 
\begin{eqnarray}
R \left( r_0=R, t \right)  &\sim&  
\frac{\alpha}{2} \left( \frac{2 \pi \alpha}{\sin \pi \alpha} \right)^{3/2} 
\frac{1}{\displaystyle \Gamma \left( 1 - \alpha/2 \right)} 
\left( \frac{\displaystyle \gamma_{\rm rc}}{\displaystyle \gamma_{\rm r}^{3/2}} \right)^{\alpha} 
\frac{1}{\displaystyle t^{1+\alpha/2}}
\nonumber \\
&\sim& \frac{\displaystyle C_1}{\displaystyle \left( \gamma_{\rm r}^{3/2} /\gamma_{\rm rc} \right)^\alpha t^{1+\alpha/2}} ,
\label{contactasym}
\end{eqnarray}
where $C_1$ is a constant. 
The intensity of photoluminescence decay depends on the relative magnitudes of $\gamma_{\rm r}$ and $\gamma_{\rm rc}$. 
The photoluminescence is weak when the jump frequency of electron migration is large 
because electrons escape recombination by diffusion. 
Strong photoluminescence is expected when the intrinsic recombination rate is large.

In the second case the initial distance between the electron and the hole of a geminate pair is assumed to be large and the intrinsic recombination rate 
is assumed to be infinitely large. 
Then we obtain from Eq. (\ref{asympt}), 
\begin{eqnarray}
R \left( \vecr_0, t \right)  
\sim \frac{\displaystyle C_2}{\displaystyle \sqrt{\displaystyle \gamma_{\rm r}}^\alpha t^{1+\alpha/2}} , 
\label{lrasym}
\end{eqnarray}
where $C_2$ is a constant \cite{Barkai2001}. 
Intensity becomes smaller as the jump frequency increases.

In both cases the photoluminescence intensity has a scaling form, 
\begin{equation}
R \left( \vecr_0, t \right)  \sim \frac{1}{\displaystyle C^\alpha t^{1+\alpha/2}} , 
\label{scaling}
\end{equation}
where $C$ is a constant. 
The photoluminescence decay measured by Murayama {\it et al. }
obeys the scaling given by Eq. (\ref{scaling}), 
if temperature dependence arises only from the disorder parameter, $\alpha$,  
and values of $\alpha$ measured by the time-of-flight experiments at each temperature are introduced. 
The temperature dependence of $\alpha$ is phenomenologically described by a linear function, 
\begin{equation}
\alpha = \alpha_0 + \frac{T}{T_0},
\label{alpha}
\end{equation}
as shown in Fig. 1, where $\alpha_0=0.15$ and $T_0= 313.4$ [K]. 
If we assume an exponential distribution of band tail states characterized by an attenuation parameter $k_{\rm B} T_0$, 
$\alpha$ is given by $\alpha=T/T_0$.
The presence of the constant term, $\alpha_0$, in Eq. (\ref{alpha}) 
implies 
that activated release of an electron from 
an exponentially distributed band tail states is a too simplified model. 
Nevertheless,  
the linear function is a good approximation for observed values of 
the disorder parameter, $\alpha$. 
By introducing the linear temperature dependence of $\alpha$, Eq. (\ref{alpha}), 
into the scaling relation, Eq. (\ref{scaling}), 
photoluminescence intensity at a certain time, $t=t_0$, after photo-excitation 
is rewritten in the following way as a function of temperature, 
\begin{equation}
R \left( \vecr_0, t_0 \right)  \sim \exp \left( - \frac{T}{T_c} \right), 
\label{Tscaling}
\end{equation}
where $T_c=T_0/\ln \left( C \sqrt{t_0} \right)$. 
In Fig. 2, 
photoluminescence observed at a delay time of 
$t_0=10$ [$\mu$sec] is plotted against temperature. 
The line represents Eq. (\ref{Tscaling}) with $T_c=30$ [K]. 
The similar temperature dependence is observed for 
quantum yield of photoluminescence  \cite{CPP,Street}. 
Since the kinetics of photo-luminescence decay is highly non-exponential, 
quantum yield does not necessarily obey 
the same temperature dependence as photoluminescence intensity 
at a certain delay time. 
Eq. (\ref{Tscaling}) is proposed for quantum yield 
by assuming activated behavior for the competition between radiative and non-radiative processes  \cite{CPP}. 
On the other hand, in our derivation of photoluminescence intensity at a certain delay time, 
escape of an electron from recombination is a non-radiative process 
which competes with radiative recombination. 
By substituting experimental values of 
$T_0$, $T_c$, and $t_0$ into the relation, 
$C \sqrt{t_0} = \exp \left( T_0 / T_c \right)$, 
we find, 
$C^2= 6.8 \times 10^{13} [1/s]$. 
Now, we consider two extreme situations leading to the same scaling form Eq. (\ref{scaling}). 
Although the temperature dependence is the same, 
different values of parameters may be assigned in these two cases. 
In the first case we find $\gamma_{\rm r} \sim 10^{14} \mbox{[1/s]}$ for $\gamma_{\rm r} \sim \gamma_{\rm rc} \mbox{[1/s]}$. 
The values are consistent with $\gamma_{\rm r} \sim 10^{14}-10^{15} \mbox{[1/s]}$ estimated  from 
the relation $(L/b)^2 \sim \left(\gamma_{\rm r} t_{\rm tr} \right)^\alpha$, 
where $\alpha \sim 0.6$, the
zero-field value of the transit time, $t_{\rm tr} \sim 10^{-4} \mbox{[s]}$, in time-of-flight experiments with the sample thickness, 
$L \sim 5 \mbox{[}\mu \mbox{m]}$, and the jump length, 
$b \sim 20-50 \mbox{\AA}$, assumed for hopping random walk of an electron are introduced.  
In the second case we obtain $\gamma_{\rm r} \sim 6.8 \times 10^{13} \mbox{[1/s]}$ 
which is also consistent with the value estimated above. 
The measured photoluminescence obeys the scaling form, 
Eq. (\ref{scaling}), 
with reasonable values of parameters for each case. 
Unfortunately, it is impossible to differentiate the results of these two extreme situations only  
from the scaling relation of long time asymptotic.

In order to see the time range 
over which the long time asymptotic, Eq. (\ref{scaling}), is valid, 
lines given by Eq. (\ref{scaling})  
are compared  in Fig. 3 with measured photoluminescence decay.  
It is difficult to keep all experimental conditions unaltered 
when temperature is changed. 
Therefore, intensity difference as a function of temperature is not reliable 
and we focus our attention on its time dependence; 
we used Eq.  (\ref{scaling}) with temperature dependence of
$\alpha$ given by Eq. (\ref{alpha}) and $C$ as a fitting parameter. 
As we can see from Fig. 3, 
the lines given by Eq. (\ref{scaling}) are consistent with experimental results at long times at relatively high temperature, 
$\mbox{T} \geq 105 \mbox{[K]}$. 
However, short time kinetics deviates from long time algebraic decay. 
The transition occurs at larger values of time for lower temperature. 
Below $100$ [K], 
the photoluminescence decay at $t_0=10$ [$\mu$sec] is 
not in the regime of long time asymptotic. 
This is the reason for the deviation from the scaling relation 
at temperatures below $100$ [K] in Fig. 2. 
At low temperatures electron-hole recombination tends to occur 
before electron executes random walk by thermal hopping. 
Such short time kinetics is out of the range of applicability of the present model 
and requires extending our theory 
to include the precise distance dependence of the intrinsic rate. 
Nevertheless, 
the observed slow decay kinetics up to 10 microseconds at $105$ [K] 
in Fig. 3 suggests 
that the intrinsic reaction rate has a finite value and competes with hopping transition. 
Thus, we can draw the following picture. 
Initially an electron is photo-excited close to a hole by band edge excitation. 
The hole is immediately trapped. 
The electron executes random walk by thermal hopping. 
It is most likely that the electron does 
not necessarily recombine with the hole even when it reaches a recombination distance 
\cite{rd}.

The recombination rate in a lattice model is also derived, where 
an electron is excited initially at a neighboring site of a hole 
and recombination takes place with a certain rate 
when an  electron and a hole occupy the same site. 
The recombination rate 
is obtained from the 
decay rate of survival probability, $N\left( \vb_j, t \right)$, 
calculated previously as, 
\begin{equation}
 N\left( \vb_j, t \right) \propto \frac{1}{\displaystyle \left(\gamma_{\rm r}^{3/2} / \gamma_{\rm rc} \right)^\alpha t^{\alpha/2}} , 
\end{equation}
which is Eq. (6.13) of ref. \cite{SekiOctober}. 
By taking the time derivative of the above equation 
the scaling relation, Eq. (\ref{contactasym}), is reproduced.

\setcounter{equation}{0}
\section{Discussion}
\vspace{0.5cm}

The photoluminescence observed at temperatures higher than $50$ K has a long time asymptotic decay 
described by $t^{-\delta}$. 
The exponent, $\delta$, is smaller than $1.5$ predicted from a geminate recombination model 
with normal diffusion of an electron \cite{HNS}. 
If one takes into account that electron hopping between localized sites in the band tail states cannot be described by 
normal diffusion but by dispersive transport due to many time scales for detrapping,  
the exponent $\delta$ is expressed in terms of disorder parameter, $\alpha$ by 
$\delta = 1+ \alpha/2$. 
The predicted value of $\delta$ is consistent 
with experimentally measured value 
when $\alpha$ obtained by time-of-flight experiments is used. 
In the time-of-flight experiments the transit current, $I$, is described by two different power law decay functions, 
$I \propto t^{- \left(1- \alpha_i \right)}$ ($t < t_{\rm tr}$) and 
$I \propto t^{- \left(1+\alpha_f \right)}$ ($t > t_{\rm tr}$),
where $t_{\rm tr}$ is called the transit time. 
The continuous time random walk model and 
fractional Fokker-Planck equation 
predict $\alpha_i = \alpha_f$ 
\cite{SM,Barkai2001}. 
However, 
measured values of $\alpha_i$ and $\alpha_f$ are slightly different \cite{TiedjePRL,Kirby}. 
When they are different we choose $\alpha= \alpha_i$ 
because photoluminescence mainly decays within the transit time 
and the origin of discrepancy between  $\alpha_i$ and $\alpha_f$ has not been fully
understood. 
A possible origin is the structured distribution of localized states 
\cite{MarshallPMB,Marshall}. 
The other possibility is that it is due to the correlation between waiting times which is neglected in the continuous time random walk model. 
In fact different values of 
$\alpha_i$ and $\alpha_f$  are obtained in the hopping simulation 
in which the correlation between hopping frequencies exists 
\cite{MurayamaMC1}.

Theory and experiment can also be compared by evaluating the temperature dependent decrease of the luminescence intensity 
 due to escape of an electron-hole pair from recombination. 
The measured intensity of photoluminescence obeys a scaling relation, Eq. (\ref{scaling}). 
The scaling relation is obtained from a geminate recombination model with dispersive diffusion. 
The reasonable parameter values are obtained by assuming that initially an electron is photo-excited close to a hole 
and the intrinsic recombination rate of an electron with a hole is not infinitely large.

The competition between radiative recombination and non-radiative escape processes of diffusion
is taken into account in geminate recombination model by introducing a sink term for radiative recombination and 
dispersive diffusion for transport. 
Recently, we have derived a fractional reaction-diffusion equation from a continuous time random walk model \cite{SekiJuly,FDASeki}. 
Recombination rate, Eq. (\ref{asympt}), can be also derived from the fractional reaction-diffusion equation, 
\begin{eqnarray}
\frac{\partial}{\partial t} \rho \left( r, t \right) = \frac{\partial}{\partial t}
\int_0^t d\,t_1 \frac{1}{\Gamma ( \alpha )} \frac{1}{\left( t - t_1 \right)^{1-\alpha}}
\left[ D_{\alpha} \nabla^2 \rho \left( r, t_1 \right) - k_{\alpha} 
\frac{\delta \left( r - R \right)}{4 \pi R^2} \rho \left( r, t_1 \right) \right] .
\label{fundamentalold}
\end{eqnarray}
Another remark on the temperature dependence of the photoluminescence intensity is that 
measured values of $\alpha$ is not proportional to temperature for some semiconductors \cite{Naito}. 
In principle, the measured values of $\alpha$ for each temperature should be employed when we compare 
theoretical results with experimental data.

As in the previous geminate recombination model for normal diffusion, 
we have neglected the coulombic interaction 
of an electron with a hole, 
although Onsager radius in these amorphous semiconductors are known to be large close to
 $100 \mbox{\AA}$
\cite{NHS,HNS}.
For normal diffusion, the power of long time algebraic decay of recombination rate 
is independent of coulombic interaction \cite{NHS,HNS}.
In the presence of distributed site energies of localized states, 
we should further take into account the relative magnitude of the coulombic interaction against 
the fluctuation of activation energy associated with hopping transition. 
When an electron is distance $r$ away from a hole 
coulombic potential is $U(r)= - e^2/(\epsilon r)$, 
where $\epsilon \sim 10$ is the dielectric constant of a-Si:H \cite{HNS}. 
Coulombic interaction creates energy difference for a hopping transition with a distance  
$b \sim 20-50 \mbox{\AA}$, 
which is estimated to be 
$b d U(r)/(dr) = e^2 b /(\epsilon r^2)$, 
while the fluctuation of activation energy associated with a hopping transition 
is characterized by the mean depth of distributed site energies of localized states, 
$k_{\rm B} T_0$ with $T_0 \sim 313$ [K]. 
By equating these two quantities, 
we find a characteristic length, 
$R_{\rm s} \equiv \sqrt{e^2 b/ \left(\epsilon K_{\rm B} T_0 \right)} 
\sim 40 \mbox{\AA}$, 
which is much smaller than the Onsager length. 
When distance between a hole and an electron 
 is larger than the characteristic length, $R_{\rm s}$, 
coulombic interaction can be neglected. 
Since $R_{\rm s}$ is small and comparable to the hopping distance 
and we are interested in the long time asymptotic of photoluminescence, 
where electron migration at large distances from a hole is important, 
the power of the long time asymptotic of recombination rate would not 
be influenced by coulombic interaction. 
The interaction may be important for the decay of photoluminescence at short times, 
which we do not investigate in this paper. 
Further theoretical study is required to clarify the influence of coulombic interaction
on recombination rate.

A probable mechanism of geminate recombination is radiative tunnelling which occurs at short distance comparable to 
hopping distance \cite{Street,NHS}. 
Such short range reaction can be modeled by recombination reaction at encounter distance 
as long as long time asymptotic is concerned. 
Distance dependence of intrinsic reaction rate matters for short time kinetics. 
The more precise model should be introduced to account for the photoluminescence decay at short times. 
Murayama {\it et al.} has developed a theoretical model for dispersive transport, 
where the energy fluctuation of the localized band tail states is self-affine \cite{Ando,MurayamaMC1,MurayamaMC2,MurayamaMC3,MurayamaMC5}. 
By the Monte Carlo simulation the photoluminescence decay 
over the whole range of time are well reproduced as well as the dispersive transit currents. 
More recently, the fractional diffusion equation is numerically solved by Fukunaga 
to reproduce the  photoluminescence decay over the whole range of time \cite{Fukunaga}, 
although his way of including recombination reaction is different from ours because the reaction term is not coupled to 
the memory kernel like Eq. (\ref{fundamentalold}) \cite{Sung,SekiJuly,FDASeki}. 
The long time asymptotic is different from ours when 
$\alpha > 2/3$ \cite{SekiOctober}. 
Although our theory is restricted only to the long time asymptotic decay,  
it provides an analytical relation between the exponent of algebraic photoluminescence decay and  
the disorder parameter $\alpha$, which can be obtained from the time-of-flight experiments. 
The amplitude of the power law decay of photoluminescence is also given in terms of the disorder parameter, $\alpha$. 
Some new experimental evidence is also presented in this paper, 
which supports our analytical results.

\acknowledgments
We are grateful to Y. Ando for her assistance in experiments. 
This work is supported by the COE development program of 
MEXT
and the Grant-in-Aid for Young Scientists(B) (14740243) from 
JSPS.


\newpage
\vskip1cm
\noindent
{\bf Figure 1:}~
The exponent $\delta$ of the long time photoluminescence decay and the disorder parameter $\alpha$, in a-Si:H against temperature. 
Closed triangles are  $\delta$ obtained  from the 
total light decay of photoluminescence in a-Si:H  excited with the pulsed laser of photon energy of $2.32$ [eV]. 
Closed circles are $\alpha$ obtained from electron time-of-flight experiments in a-Si:H \cite{MurayamaMC5}. 
Open circles are calculated by $\delta=1+\alpha/2$ from values of $\alpha$ shown by closed circles.
The solid line is a linear fit, $\alpha= 0.15+T/313.4$.
\vskip1cm
\noindent
{\bf Figure 2:}~
Temperature dependence of photoluminescence intensity from 
a-Si:H excited with the pulsed laser of photon energy of $2.32$ [eV]. 
The photoluminescence has been observed at the delay time $10 $ [$\mu$sec] 
and the peak energy of $1.24$ [eV]. 
Solid lines is $\exp \left( - T /T_c \right)$ with $T_c=30$ [K]. 
\vskip1cm
\noindent
{\bf Figure 3:}~
Total light decay of photoluminescence in a-Si:H  excited with the pulsed laser of photon energy of $2.32$ [eV]. 
The photoluminescence has been observed at different temperatures, 105[K] , 
145[K] and 170[K] (top to bottom) \cite{FDAMurayama}. 
Solid lines are theoretical lines given by Eq. (\ref{scaling}). 
\newpage
\begin{flushright}
Figure 1
\end{flushright}
\begin{figure}[h]
\includegraphics[width=10cm,clip]{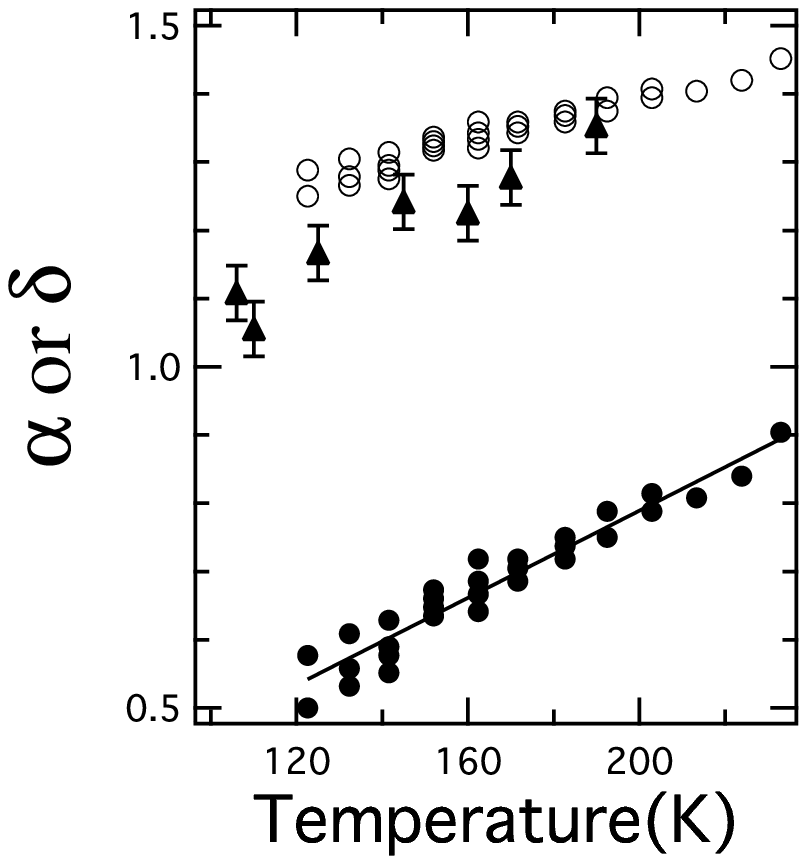}
\end{figure}
\newpage
\begin{flushright}
Figure 2
\end{flushright}
\mbox{ }\\
\begin{figure}[h]
\includegraphics[width=10cm]{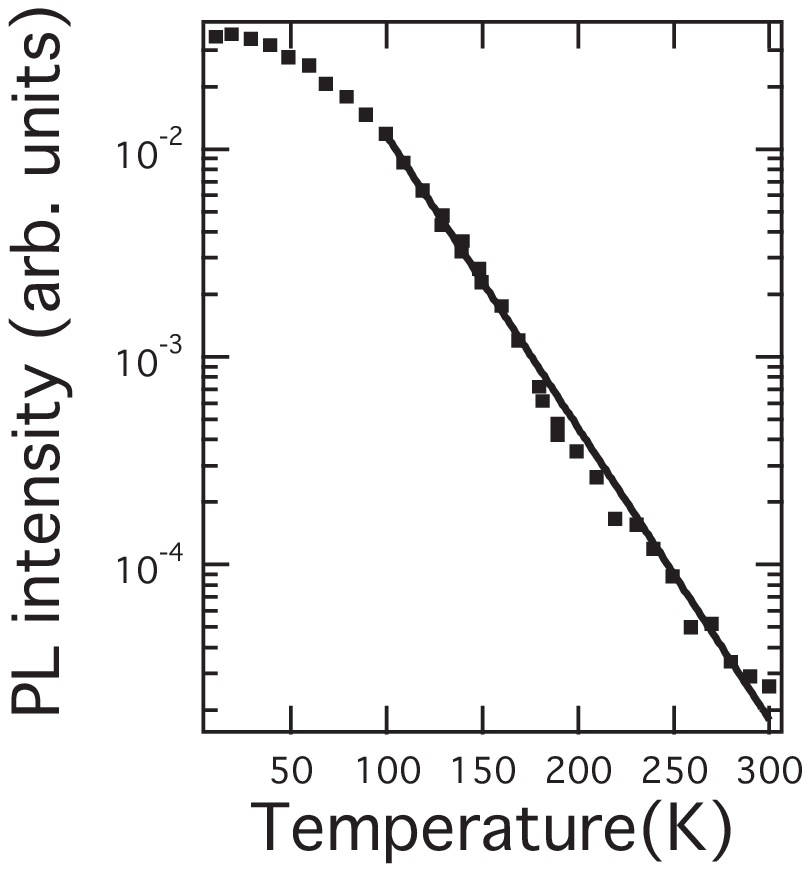}
\end{figure}
\newpage
\begin{flushright}
Figure 3
\end{flushright}
\mbox{ }\\
\begin{figure}[h]
\includegraphics[width=10cm]{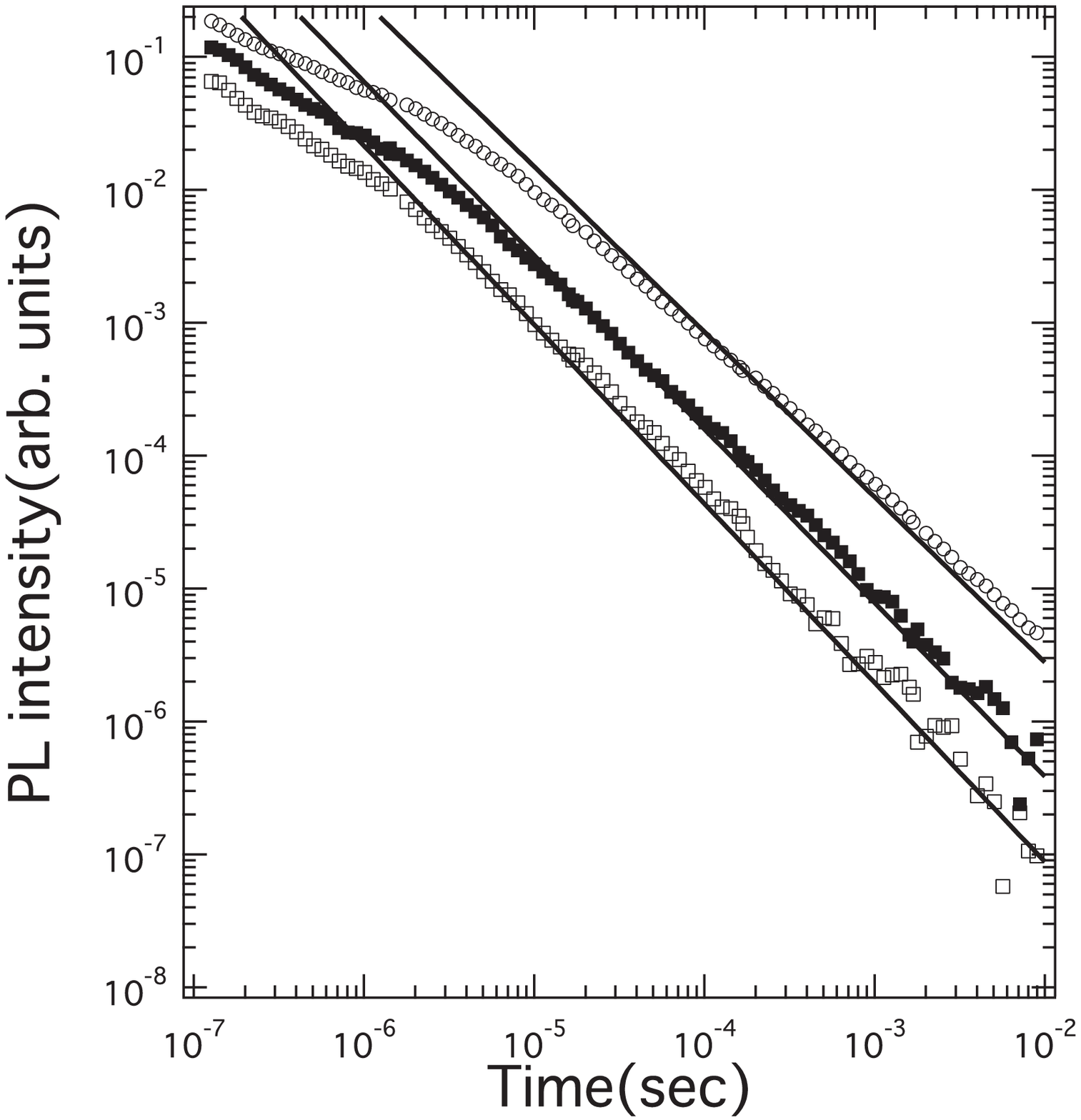}
\end{figure}

\begin{references}
\bibitem{rd}  
The recombination distance is the distance between an electron and a hole at which they recombine. If the intrinsic recombination rate at the recombination distance is infinite, an electron and a hole always recombine, once they come within the recombination distance. However, the intrinsic recombination rate at the recombination distance is in general finite. Therefore, even if an electron and a hole come within the recombination distance, some fraction of the pair will escape recombination and separate away again.
\end{references}
\end{document}